\documentclass[conference, 10pt]{IEEEtran}
\ifCLASSINFOpdf
\else
\fi
\usepackage{graphicx}
\usepackage{color}
\usepackage{placeins}
\usepackage{float}
\usepackage{tabularx,colortbl}
\usepackage{subfig}
\usepackage{epsfig}
\usepackage{epstopdf}

\begin{document}
\title{AV1 Video Coding Using Texture Analysis \\ With Convolutional Neural Networks}

\author{\IEEEauthorblockN{Di Chen, Chichen Fu, Fengqing Zhu}
\IEEEauthorblockA{School of Electrical and Computer Engineering\\
Purdue University\\
West Lafayette, Indiana, USA\\}
\and
\IEEEauthorblockN{Zoe Liu}
\IEEEauthorblockA{Google, Inc.\\
Mountain View, Califronia, USA}}


%Telephone: (800) 555--1212, Fax: (888) 555--1212}


\maketitle

\begin{abstract}
%\boldmath
Modern video codecs including the newly developed AOM/AV1 utilize hybrid coding techniques to remove spatial and temporal redundancy. However, efficient exploitation of statistical dependencies measured by a mean squared error (MSE) does not always produce the best psychovisual result. One interesting approach is to only encode visually relevant information and use a different coding method for ``perceptually insignificant" regions in the frame, which can lead to substantial data rate reductions while maintaining visual quality. In this paper, we introduce a texture analyzer before encoding the input sequences to identify detail irrelevant texture regions in the frame using convolutional neural networks. We designed and developed a new coding tool referred to as texture mode for AV1, where if texture mode is selected at the encoder, no inter-frame prediction is performed for the identified texture regions. Instead, displacement of the entire region is modeled by just one set of motion parameters. Therefore, only the model parameters are transmitted to the decoder for reconstructing the texture regions. Non-texture regions in the frame are coded conventionally. We show that for many standard test sets, the proposed method achieved significant data rate reductions.
\end{abstract}


% \fi
\IEEEpeerreviewmaketitle

\section{Introduction}
The Alliance for Open Media (AOM) \cite{AOM} is a joint effort between Google and several other industrial leaders, set to define and develop media codecs, media formats, and related technologies that is open-source and loyalty-free to meet the expanding need in web-based video consumption. We propose a new coding paradigm that leverages techniques of texture analysis and synthesis to achieve coding gains and contribute to the first edition of the AOM video codec, namely AV1 \cite{av1-joshi2017,av1-liu2017,chen2018,fu2018}.

Modern video codecs utilize hybrid coding techniques consisting of 2D transforms and motion compensation techniques to remove spatial and temporal redundancy. Our approach is different in that we will only encode, using AV1, areas of a video frame that are ``perceptually significant." The ``perceptually insignificant" regions will not be encoded. By ``perceptually insignificant" pixels we mean regions in the frame that an observer will not notice any difference without observing the original video sequence. The encoder fits a model to the perceptually insignificant pixels in the frame and transmits the model parameters to the decoder as side information. The encoder uses the model to reconstruct the pixels. This is referred to as the ``analysis/synthesis" coding approach. The use of texture segmentation and synthesis approach to reconstruct texture region with acceptable perceptual quality for still images were proposed in some earlier works \cite{peterson1990,kunt1985,delp1979}. We extended similar ideas to video coding in our previous work \cite{bosch2011}, where we developed a feature based texture analyzer to identify perceptually insignificant regions in the frame and classify them into texture classes. At the encoder, instead of performing inter-frame prediction to reconstruct these regions, displacement of the entire texture region is modeled by a set of motion parameters. The motion parameters and the texture region information are coded and transmitted separately as side information. We have shown that data rate reductions of 5-20\% can be achieved using this approach when implemented in H.264.

While the feature based texture analyzer requires a proper set of parameters to achieve accurate texture segmentation for different videos, deep learning based methods usually do not require such parameter tuning for inference. Recently, deep learning based methods have been developed and applied to different aspects of video coding and has shown promising performance. In \cite{park2016}, a new in-loop filtering technique using convolutional neural network called IFCNN is presented. It outperforms the conventional in-loop filtering method done by a de-blocking filter followed by sample adaptive offset filter (SAO) with respect to coding efficiency and subjective visual quality in HEVC. In \cite{wang2017}, a very deep convolutional network, DCAD, is developed to automatically remove the artifacts and enhance the details of HEVC-compressed videos at the decoder end. DCAD improves the visual quality of the reconstructed frame by automatically learn a non-linear mapping from the decoded frame to an artifact-free reconstruction.  In our previous work \cite{fu2018}, we proposed a block-based texture segmentation method to extract texture regions in a video frame using convolutional neural networks. 

The problem with using the texture analyzer alone to encode the texture region in the video is that if each frame is encoded separately, the areas that have been reconstructed with the texture models will be obvious when the video is displayed. This then requires that the textures to be modeled both spatially and temporally. In this paper, we propose a new AV1 coding paradigm that utilizes the texture segmentation result from \cite{fu2018} by introducing a new coding mode - texture mode. The texture mode is completely an encoder side option, which in essence skips the coding of the block entirely through leveraging the use of global motions provided by the AV1 baseline. Specifically, the texture mode uses a modified version of the global motion coding tool in the AV1 codec \cite{global_motion} to ensure temporal consistency of the texture regions between frames. Based on the selection of coding structures and choices of reference frames, we investigate three different implementations of the texture mode in terms of data rate savings and perceived quality. Experimental results validate the efficacy of the texture mode with a consistent coding gain compared to the AV1 baseline over a variety of video test sets given a fixed perceptual quality level.
\section{Texture-Base Video Coding}
\begin{figure}[ht]
\includegraphics[width=1\columnwidth]{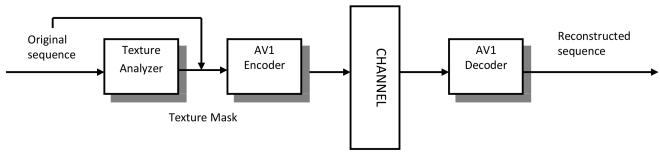}
\caption{Overview of Texture-Based Video Coding}
\label{fig:blockdia}
\end{figure}

The general scheme for video coding using texture analysis and synthesis is illustrated in Figure \ref{fig:blockdia}. The texture analyzer identifies homogeneous regions in a frame and labels them as texture. We use a classification convolutional neural network to label each block in a frame as textures or non-texture and generate a block-based texture mask for each frame. The texture mask and the original frame are passed into the AV1 video codec to enable the texture mode where the identified texture regions skip the encoding process. The texture region is synthesized by warping texture region in a reference frame to the current frame. A modified version of the global motion tool in AV1 is used to obtain motion estimation and synthesize the texture region without sending residues for the identified texture region.
%Figure \ref{fig:blockdia} shows the block diagram of our system. AOM/AV1 video codec uses the texture information extracted from a convolutional neural network texture detector to skip the encoding process on the identified texture regions. The texture region is synthesized by warping texture region in a reference frame to the current frame. A built-in global motion tool in AOM/AV1 is used to obtain motion estimation and synthesize the texture region without sending residues for identified texture region.

\subsection{Texture Analysis Using CNN}
\begin{figure}[ht]
\includegraphics[width=1\columnwidth]{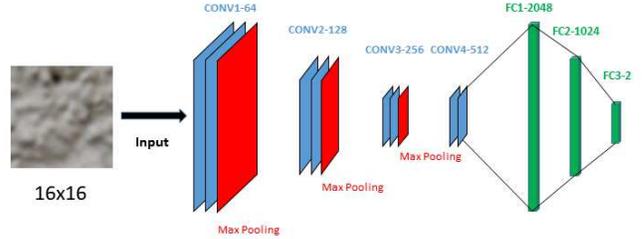}
\caption{CNN Architecture for Block-Based Texture Classification}
\label{fig:archit}
\end{figure}

A block-based segmentation method \cite{fu2018} is used to identify detail irrelevant texture regions in each frame which are potential candidates to use the texture mode in AV1. We designed a classification convolutional neural network inspired by the VGG network architecture \cite{Simonyan2014} to label a block as texture or non-texture (Figure \ref{fig:archit}). The input of our network is a $32\times32$ color image block. The output is the probability that the image block contains texture or non-texture.

In \cite{fu2018}, image patches of size $16\times16$ with texture and non-texture labels are used to train the network. The image patch size was increased to $32\times32$ to avoid detecting small moving objects in our training. Texture and non-texture images are obtained from the Salzburg Texture Image Database (STex) \cite{stex} and Places365 \cite{place365}. STex contains images with single texture type and images in Place365 are nature scenes with multiple objects. To create multi-resolution training examples for texture classes, images from STex are cropped from $512\times512$ into $256\times256$ and $128\times128$, followed by downsampling them to $32\times32$ . Since a texture region with consistent content are desired, images contains multiple objects should be classified as non-texture class. Therefore, images from Place365 are directly downsampled to $32\times32$  image patches to create non-texture examples that contain multiple objects.

This method was implemented in Torch \cite{torch}. A stochastic gradient descent (SGD) with momentum is used to train our network. A learning rate of 0.01, a momentum of 0.9 and weight decay of 0.00005 were used in our training. A set of training data with 1740 texture examples and 36148 non-texture examples were used to train our network. A binary cross entropy loss was used as the loss function. Since our training set are highly unbalanced, the weights of each class in the binary cross entropy loss function were set proportion to the inverse of the class frequency. A total 100 epochs were trained using mini batch size of 512 on one NVIDIA GTX TITAN GPU.

After training the CNN, texture segmentation is performed on each test video frame. Each frame is divided into $32\times32$ adjacent non-overlapping blocks. Each block in the video frames is classified as either texture or non-texture. The segmentation mask for each frame is formed by grouping the classified blocks in the frame.

\subsection{A New AV1 Coding Tool - Texture Mode}
In this section, we describe how we modified the AV1 codec by introducing a texture mode to encode the texture blocks.

\subsubsection{Texture Mode Encoder Design}
\begin{figure}[t]
\begin{center}
\noindent
  \includegraphics[width=\columnwidth]{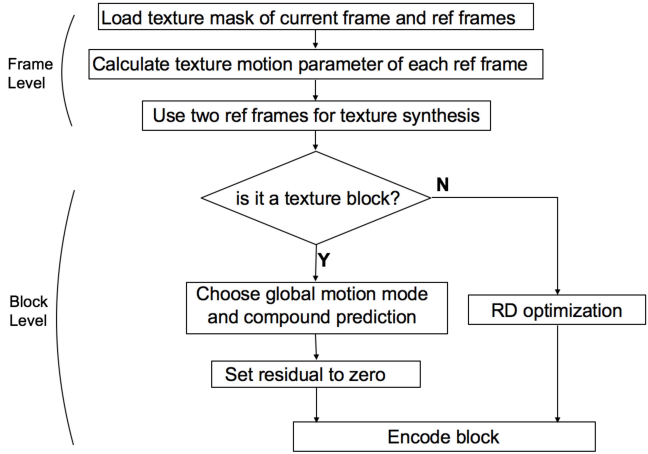}
  \caption{Texture Mode Encoder Implementation}
  \label{Fig1Label}
\end{center}
\end{figure}

The texture analyzer is integrated into the AV1 encoder as illustrated in Figure \ref{Fig1Label}. At the encoder, for each frame that contains texture area, we first fetch the texture masks for the current frame and the two corresponding reference frames from the texture analyzer. Based on the texture region in the current frame, a set of texture motion parameter that represents the global motion of the texture area is estimated for each reference frame. Then for each block larger than $16 \times 16$, we use a two-step method to check if a block is a texture block. A texture block is reconstructed using texture synthesis method thus no motion compensation residuals will be coded and transmitted for the texture block. We call this new coding paradigm the \textit{texture mode}. At the decoder, since there is no syntax change to the AV1 bitstream, the bitstream is decoded the same as AV1 baseline.

\begin{table*}[ht]
\begin{center}
\caption{Configuration of Different Texture Mode Implementation}
\label{table:config}
\begin{tabular}{|l|l|l|}
\hline
\textit{tex-all}                                                     & \textit{tex-sp}                                         & \textit{tex-cp}                                   \\ \hline
\multicolumn{3}{|l|}{Disable texture mode for GOLDEN / ALTREF frame}                                                                                     \\ \hline
Original GF group interval (4-16)                           & \multicolumn{2}{l|}{Fixed 8 GF group interval}                                             \\ \hline
single-layer coding structure                               & \multicolumn{2}{l|}{multi-layer coding structure}                                          \\ \hline
Use texture mode for all frames except GOLDEN /ALTREF frame & \multicolumn{2}{l|}{Use texture mode for every other frame (frame1,3,5,7 in the GF group)} \\ \hline
Use single-prediction (forward or backward)                 & Use single forward prediction                   & Use compound prediction                  \\ \hline
\end{tabular}
\end{center}
\end{table*}

\begin{figure}[htp]
\subfloat[GF Group Coding Structure Using \textit{tex-all} Configuration]{%
  \includegraphics[width=\columnwidth]{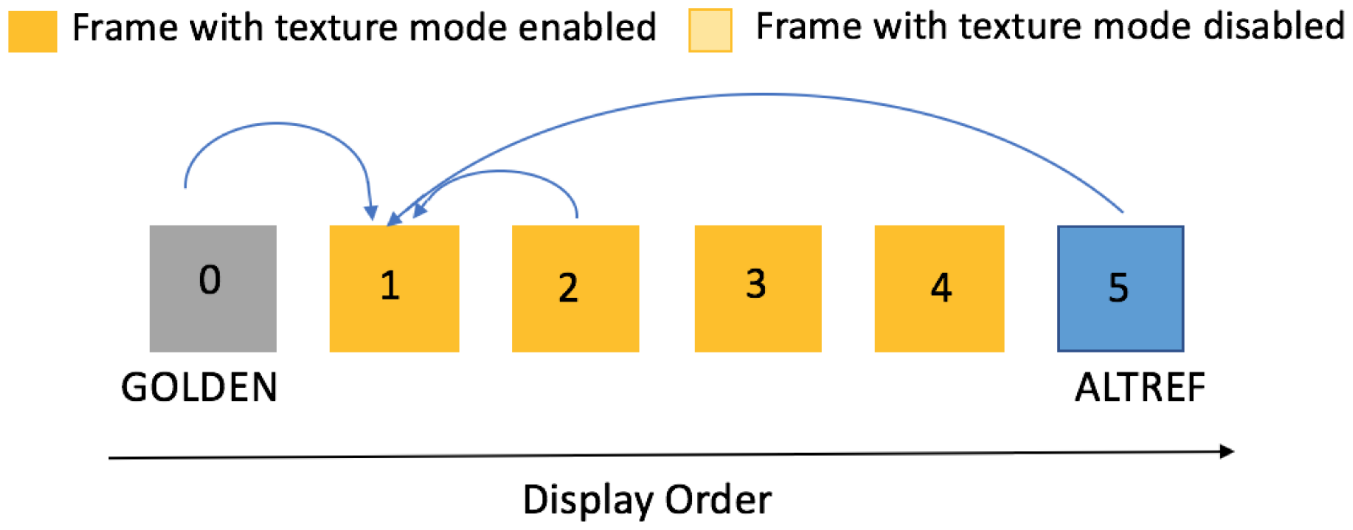}%
}

\subfloat[GF Group Coding Structure Using \textit{tex-sp} Configuration]{%
  \includegraphics[width=\columnwidth]{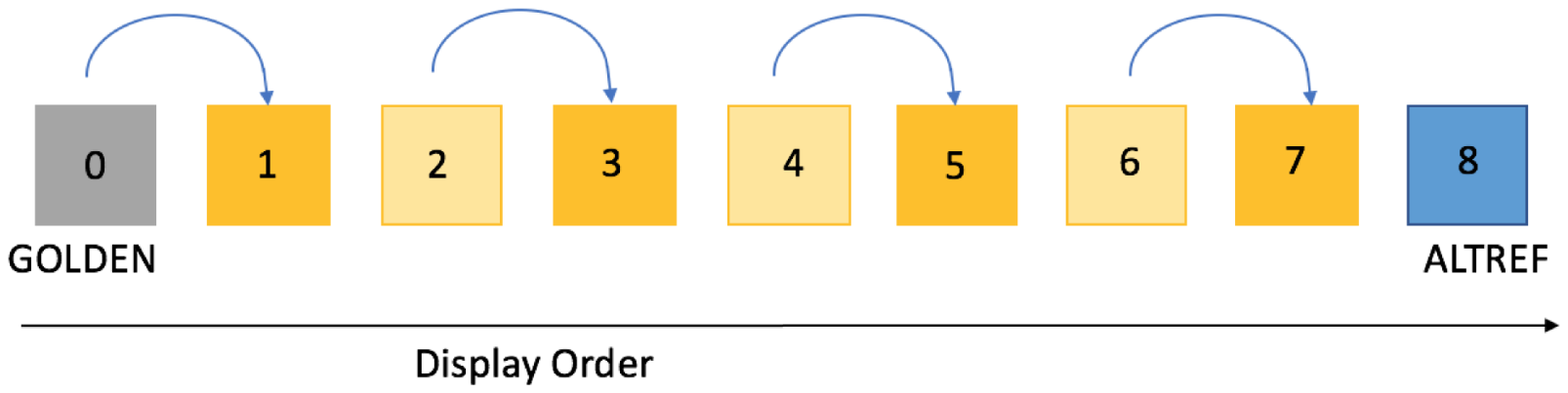}%
}

\subfloat[GF Group Coding Structure Using \textit{tex-cp} Configuration]{%
  \includegraphics[width=\columnwidth]{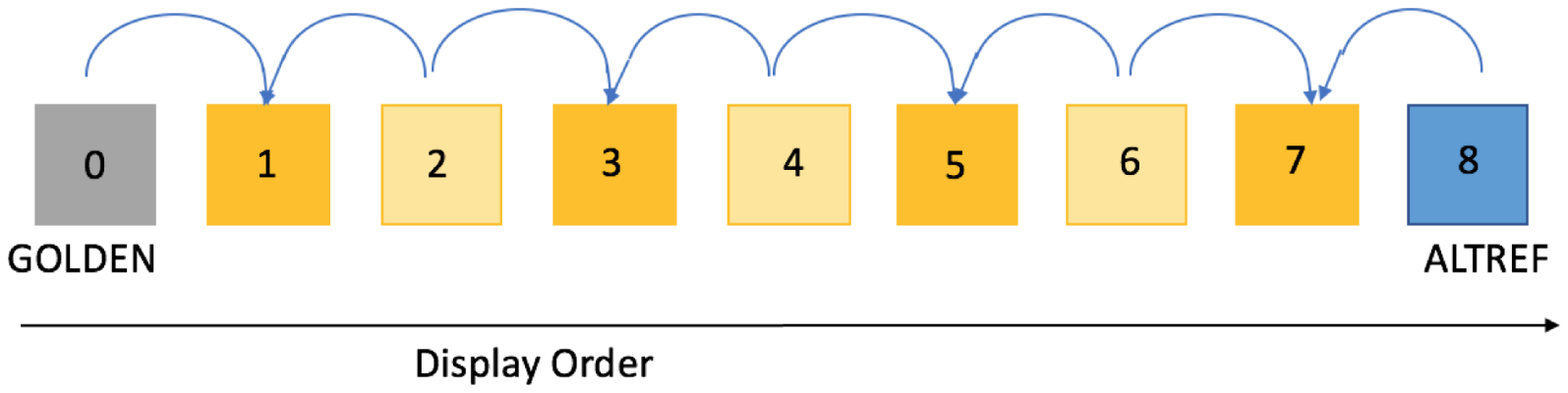}%
}

\caption{Coding Structure of Texture Mode}
\label{gf_config}
\end{figure}

In general, a texture block in the current frame is reconstructed by warping the texture block from the reference frame towards the current frame. We use a modified version of the global motion coding tool \cite{global_motion} in the AV1 codec to perform block warping as described in Section \ref{sssec:motion}. 
Based on the selection of coding structures and choices of reference frames for texture synthesis, we investigated three different implementations, namely \textit{tex-all}, \textit{tex-sp}, and \textit{tex-cp} of the texture mode in terms of data rate savings and perceived quality. Configuration of the three implementations are described in Table \ref{table:config} and can be visualized in Figure \ref{gf_config}. For \textit{tex-sp} and \textit{tex-cp}, a multi-layer coding structure \cite{multilayer} is used for each GF group.

The \textit{tex-all} implementation has the best data rate savings since the number of frames with texture mode enabled is approximately twice as many as the other two implementations. However, we observed visual artifacts in its reconstructed videos in several test sequences due to the accumulated error from warping displacement. The artifacts are most prominent in videos with high motion or complex global motion.

The \textit{tex-sp} implementation solves the accumulation error by only enabling texture mode for every other frame. It only uses the immediate previous frame as the reference frame for texture warping to get more accurate global motion model. As a result, the data rate savings are reduced to approximately half the data rate savings of  the \textit{tex-all} configuration. Some flickering artifacts can still be observed between frames for some test videos.

The \textit{tex-cp} further reduces the flickering artifacts by using compound prediction from the previous frame and the next frame. The data rate savings are only slightly lower than that of \textit{tex-sp}. The improvement in visual quality is most obviously in low-mid resolution videos. Therefore, we select \textit{tex-cp} to be our final configuration for the texture mode implementation.

\subsubsection{Texture Motion Parameters}
\label{sssec:motion}
The global motion coding tool in AV1 is used primarily to handle camera motion. A motion model is explicitly conveyed at the frame level for the motion between a current frame and any one or more of its reference frames. The motion model can be applied to any block in the current frame to generate a predictor. Either the planar projective or affine transformation is selected as the motion model. The motion model is estimated using a FAST feature \cite{FAST} matching scheme followed by robust model fitting using RANSAC \cite{RANSAC}. The estimated global motion parameter is added to the compressed header of each inter-frame. 

The motion model parameters of the global motion coding tool in AV1 is estimated at the frame level between the current frame and the reference frame. These parameters may not accurately reflect the motion model for the texture regions within a frame. We modified the global motion tool to deign a new set of motion modal parameters, called \textit{texture motion parameters}. The texture motion parameters is estimated based on the texture region of the current frame and the reference frame using the same feature extraction and model fitting method as in the global motion coding tool. A more accurate motion model for texture region may reduce the artifacts on the block edges between the texture blocks and non-texture blocks in the reconstructed video. In order to keep the syntax of AV1 bitstream consistent, the texture motion parameters are sent to the decoder in the compressed header of the inter frames by replacing the global motion parameters of the reference frames. Since most texture regions reside in the background, there is no significant influence on the non-texture blocks which are coded using global motion mode by replacing the global motion parameters with the texture motion parameters. 

%Since non-texture regions in the current frame are conventionally coded, it may use the global motion parameters if the global motion mode is enabled. By using instead the texture motion parameters, there could be large residual for the non-texture blocks which use global motion mode. However, in practice, texture regions usually reside in the background of a frame. 

%Because the motion model based on the texture region may not be accurate for blocks in the non-texture region, it may produces larger residuals for the non-texture blocks under global motion mode. The texture region is usually on the background of a frame. So the texture motion model and the global motion model is usually very close and its influence on the non-texture blocks is not significant.

\subsubsection{Texture Block Decision}
For the current implementation, the minimum size of a texture block is $16 \times 16$. For all blocks larger than or equal to $16 \times 16$, we use a two-step approach to check if a block should be encoded using the novel texture mode scheme we proposed to AV1. First, we overlap the texture mask generated by the texture analyzer and the current frame to check if the entire block is inside the texture region of the current frame. We also need to ensure that the pixels used for texture synthesis in the reference frames are within the texture regions identified by the texture analyzer. In the second step, we warp the blocks inside the texture region towards the two reference frames, i.e., the previous frame and the next frame in the \textit{tex-cp} configuration. If the two warped blocks are within the texture regions of both corresponding reference frames, the block is considered a texture block and texture mode is enabled. 

\subsubsection{Block Splitting Decision}
As for block splitting decision, the position of the texture regions inside of a macroblock has higher priority than the RD values of different block splitting methods for this macroblock. If the block is a texture block, we do not further split it into smaller sub-blocks. If the block contains no texture region, RD optimization is performed for block partitioning and mode decision. If part of a macroblock contains texture region, we split it into sub-blocks regardless of the RD value. In general, there is no block that is part texture and part non-texture. The use of texture mode also largely reduces the encoding speed, since no RD optimization is performed for a texture block which reduces the need for different prediction modes, reference frames selection, and block splitting recursion.

\subsubsection{Texture Synthesis}
We use AV1 codec's global motion tool and compound prediction to synthesize texture for texture blocks at the decoder. The previous frame and the next frame of the current frame are chosen to be the reference frames for the texture block reconstruction. The texture region in the two reference frames are warped towards the texture blocks in the current frame using the corresponding texture motion parameters. We used compound prediction to synthesize the texture block from the two reference frames. As discussed earlier, the use of compound prediction for texture blocks reduces flickering artifacts between frames. The residual of the texture blocks is set to zero. Since all texture blocks in one frame use the same reference frames, there are no blocky artifacts from texture synthesis on the block edges of the texture blocks within the texture region.  

\section{Experimental Results}

\subsection{Texture Analysis}
\begin{figure}[ht]
\includegraphics[width=0.9\columnwidth]{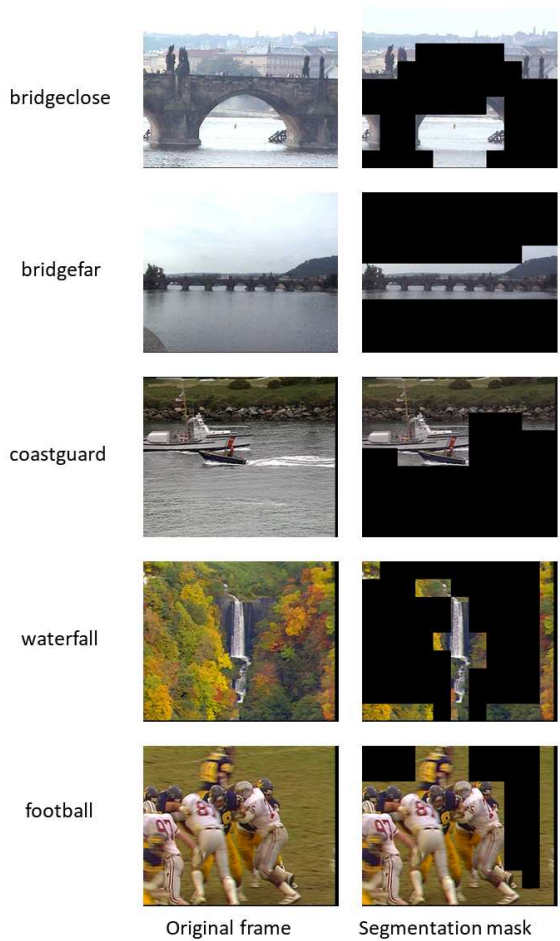}
\vspace{-0.2in}
\caption{Texture Segmentation Examples}
\label{fig:tex_res}
\end{figure}

Nine different video sequences were tested using the CNN based texture analyzer. Sample texture segmentation results are shown in Figure \ref{fig:tex_res}. Our texture analyzer successfully segments out most texture regions. Currently, the texture analyzer uses a block-based texture classification method with fixed block size. So the segmentation mask does not contain texture regions at finer levels. As a result, some of the texture regions are miss detected because one image block may contain different types of textures.

\subsection{Coding Performance}
To evaluate the performance of the proposed texture-based method, data rate savings at four quantization levels (QP = 16, 24, 32, 40) are calculated for each test sequence using the \textit{tex-cp} configuration and compared to the AV1 baseline. The AV1 baseline is the original codec with fixed golden frame and a group interval of eight frames. Data rate is computed by dividing the output WebM file size by the number of frames. The WebM file is the output bitstream from the AV1 encoder. Results for several test videos are shown in Table \ref{gain}. We also include the average percentage of pixels that uses the texture mode in a frame in the table.  
  
\begin{table}[t]
  \caption{Data Rate Savings for Different QP Level}
  \label{gain}
  \includegraphics[width=\columnwidth]{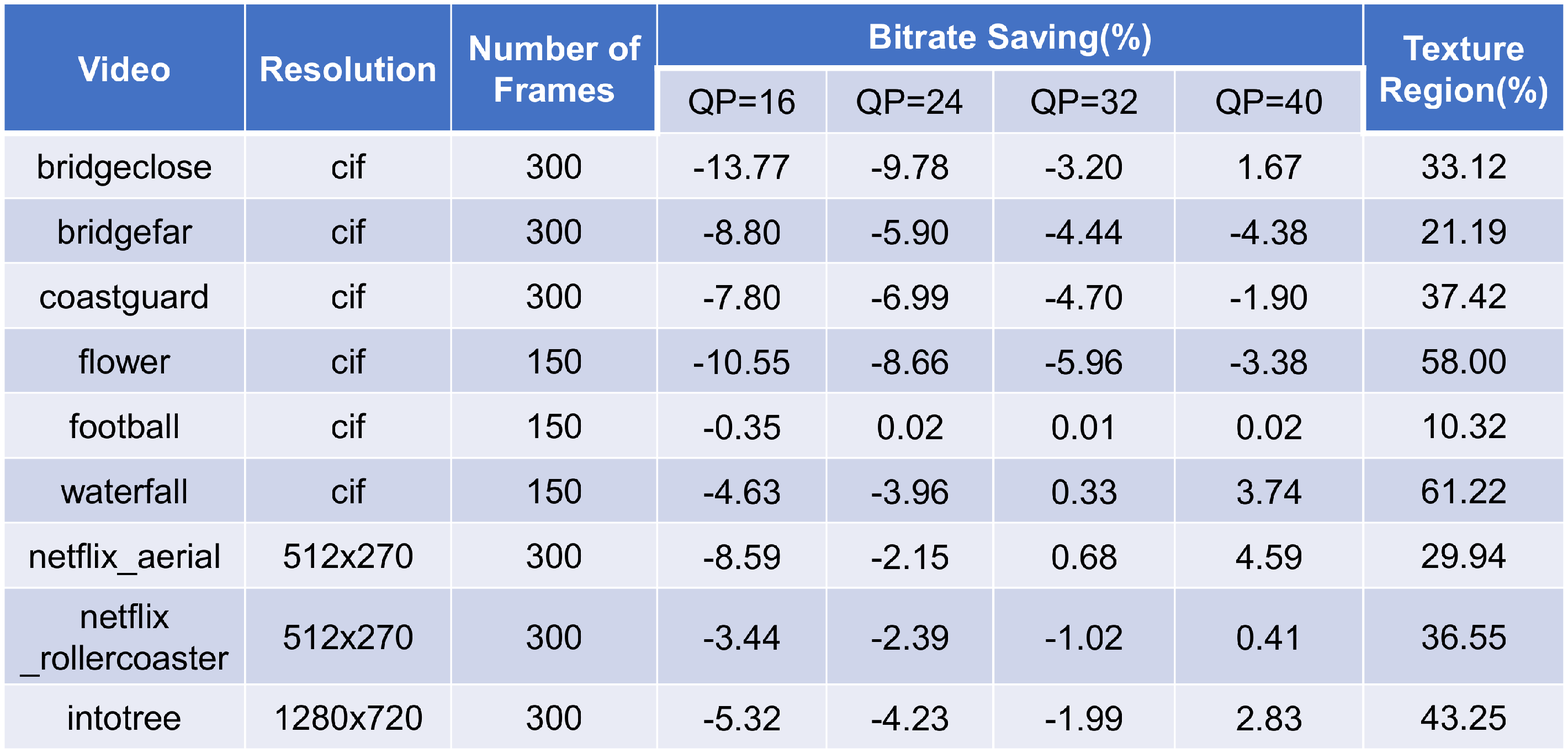}
\end{table}

As shown in Table \ref{gain}, with low QP, most of the videos show large data rate savings. However, as the QP increases, the data rate savings decreases. Some test videos, such as football, waterfall and netflix\_aerial show worse coding performance than the AV1 baseline at high QP. This is because with high QP, many non-texture blocks also have zero residual and our method requires a few extra bits for the texture motion parameters and for using two reference frames in compound prediction. 
%We also observed minor motion artifacts between the edge of texture regions and non-texture regions. Inaccurate texture motion parameters could lead to motion artifacts in our cases. However, this artifact is negligible for videos with lower resolution.   
%\subsection{Result for Preliminary Subjective Visual Quality Test}
%We performed a preliminary subjective visual quality test on five subjects. In the test, the subjects are asked to watch two versions of each test video. One is the reconstructed video using original AOM/AV1 codec with QP=15. The other is the reconstructed video using our proposed method (tex-cp) with QP=15. Then they are asked to choose the video that has better visual quality. The result of this subjective visual quality test is summarized in Table \ref{subjective}. Although he number of the subjects is statistically not enough in this preliminary subjective test, we can show that the visual quality of the reconstructed videos using our proposed method is satisfactory.

%\begin{table}
%  \caption{Subjective visual quality test result}
%  \label{subjective}
%  \includegraphics[width=\columnwidth]{./figure/subjective.eps}
%\end{table}
\section{Conclusion and Future Work}
In this paper, we proposed a new AV1 video coding paradigm that integrates a texture segmentation method into the AV1 codec. The texture segmentation method uses a deep learning based approach to detect the texture regions in a frame that is perceptually insignificant to the human visual system. A novel texture mode paradigm is proposed for an AV1 encoder, which uses the multi-layer coding structure, a modified global motion tool and the compound prediction mode. Our results showed significant increase in terms of coding efficiency compared to the AV1 baseline for a set of videos contain large texture regions. 
For our next step, we would like to assess how well the proposed texture-based method performs for different types of motion, and adjust the complexity of the motion models used depending on the texture content. 
\bibliographystyle{ieeetran}
\bibliography{egbib}
%}

\end{document}